\documentclass[12pt,a4paper]{article}
\usepackage{a4wide}
\usepackage[]{amsfonts,amsmath}
\usepackage{mathrsfs,amssymb,amscd,amsthm}

\numberwithin{equation}{section}

\newcommand{\inte}[1]{\int_{-\infty #1}^{\infty #1}\limits}
\newcommand{\intel}[1]{\int_{-\infty}^{\infty}\limits}
\newcommand{\imagunit}{\mathrm{i}}
\newcommand{\df}{\, \mathrm{d}} 
\newcommand{\ez}{\mathrm{e}}
\newcommand{\partiell}[1]{\frac{\partial}{\partial{#1}}}
\newcommand{\psl}{\mathrm{PSL}(2,\mathbb{R})}

\newtheorem{satz}{Proposition}
\newtheorem{thm}{Theorem}
\theoremstyle{definition}
\newtheorem{defn}{Definition}
\theoremstyle{plain}
\newtheorem{lemma}{Lemma}

\begin{document}

\thispagestyle{empty}

\noindent ULM-TP/06-1\\July 2006
\vspace{2cm}

\begin{center}

{\LARGE\bf  The Selberg trace formula}\\
\vspace*{5mm}
{\LARGE\bf for Dirac operators}\\
\vspace*{2cm}
{\large Jens Bolte}%
\footnote{E-mail address: {\tt jens.bolte@uni-ulm.de}}
{\large and Hans-Michael Stiepan}%
\footnote{E-mail address: {\tt hans-michael.stiepan@uni-ulm.de}}

\vspace*{1cm}

Abteilung Theoretische Physik\\
Universit\"at Ulm, Albert-Einstein-Allee 11\\
D-89069 Ulm, Germany 
\end{center}

\vfill

\begin{abstract}
We examine spectra of Dirac operators on compact hyperbolic surfaces. 
Particular attention is devoted to symmetry considerations, leading
to non-trivial multiplicities of eigenvalues. The relation to spectra 
of Maa\ss -Laplace operators is also exploited. Our main result is a Selberg
trace formula for Dirac operators on hyperbolic surfaces.
\end{abstract}

\newpage

\section{Introduction}
\label{intro}
Trace formulae play a prominent role in spectral geometry and in quantum
chaos. In spectral geometry they relate spectra of certain 
(pseudo-)~differential operators on manifolds to the geometry of that 
manifold. The prime example of such a trace formula, the classical Selberg 
trace formula \cite{Selberg:1956}, is concerned with the Laplace-Beltrami 
operator on a hyperbolic manifold, i.e., a Riemannian manifold with constant 
negative sectional curvatures. This trace formula establishes a connection 
between the spectrum of the Laplacian and the length spectrum of closed 
geodesics on the manifold.

In a semiclassical context spectra of semiclassical operators are related
to the periodic orbits of an associated Hamiltonian flow. In extensions
of this procedure to operators acting on sections in vector bundles over
the relevant phase space, it turned out that the Hamiltonian flow does
not provide the entire input on the classical side of the trace formula
(see, e.g., \cite{Bolte:1999}). In particular, for the Dirac operator
on $\mathbb{R}^3$ the corresponding classical dynamics are given in terms
of skew-product flows over Hamiltonian flows, with spin precessions as 
their cocycles. A similar result is found in the case of a trace formula
for the Dirac operator on a graph \cite{Bolte:2003}. Very recently, in the 
context of quantum ergodicity Jakobson and Strohmaier \cite{Jakobson:2006} 
observed that (the square of) a Dirac operator on a compact Riemannian
manifold is related to the associated frame flow, which is a geometric 
analogue of the skew-product flows in the semiclassical setting.

Here we introduce Dirac operators on compact hyperbolic surfaces,
investigate their spectra in terms of symmetries, and finally develop
the related Selberg trace formula in a classical setting. This is a direct
approach employing Green's functions and point-pair invariants, and
therefore complements previous work \cite{Bunke:1995}, where trace formulae 
for positive square-roots of squared Dirac operators were determined using 
Lie-algebraic methods. On the way we exploit the well-known relation of the 
squared Dirac operator to a Laplacian that contains a coupling to a magnetic 
field (see, e.g., \cite{Pnueli:1994}). In particular, we show that spectra 
of these Maa\ss -Laplacians, when they correspond to odd weights, possess 
multiplicities of at least two. This is a consequence of Kramers' degeneracy 
in the spectrum of the Dirac operator. The geometric side of the trace 
formula for the Dirac operator is primarily determined by the closed 
geodesics on the surface. In addition, we identify traces of the frame 
flow and its lift into the spin structure, which provide suitable phase 
factors associated with the closed geodesics. 

The paper is organised as follows: In Section~\ref{section:dirac} we 
review the definition of a Dirac operator on a Riemannian manifold and
explicitly carry out the constructions in the case of the hyperbolic plane. 
Section~\ref{section:auto} then is devoted to an identification of spinor
bundles over compact hyperbolic surfaces via factors of automorphy. We 
then investigate symmetries of the Dirac operator on a compact surface 
and study their influence on the spectrum in Section~\ref{section:sym}.
After having introduced point-pair invariants in 
Section~\ref{section:point-pair} we calculate traces of Hilbert-Schmidt
operators to arrive at the desired trace formula in 
Section~\ref{section:trace}. Finally, in Section~\ref{section:applications},
we discuss some applications.
\section{The Dirac operator}
\label{section:dirac}
A Dirac operator is a first order, elliptic differential operator acting 
on sections in a spinor bundle over a Riemannian manifold. For its
construction one first needs to supply an orientable, $n$-dimensional 
Riemannian manifold $(M,\mathbf{g})$ with a spin structure. To this
end, let $FM$ be the $\mathrm{SO}(n)$-principal bundle of oriented 
orthonormal frames of $TM$. A $\mathrm{Spin}(n)$-principal bundle 
$Q$ is called spin structure on $M$ if there exists a principal-bundle 
morphism $\phi:Q\rightarrow FM$ that is equivariant with respect to the 
two-fold covering $\Lambda:\mathrm{Spin}(n)\rightarrow \mathrm{SO}(n)$
(see e.g. \cite{Friedrich:1997,Baum:1981}). Moreover, in even dimensions 
$n$ the complexified Clifford algebra $\mathfrak{C}^c_n$ is isomorphic to 
$\mathrm{M}(2^{n/2},\mathbb{C})$, i.e., the algebra of complex 
$2^{n/2}\times 2^{n/2}$ matrices. The action of 
$\mathrm{M}(2^{n/2},\mathbb{C})$ on $\mathbb{C}^{2^{n/2}}$ defines the 
Clifford module $\Delta_n$. As $\mathrm{Spin}(n)$ is contained in 
$\mathfrak{C}^c_n$, $\mathrm{\Delta}_n$ is also a module for the spin group. 
This yields the spin representation 
$\rho:\mathrm{Spin}(n)\rightarrow\mathrm{Aut}(\Delta_n)$, which can be used 
to associate the spinor bundle $S:=Q\times_\rho \Delta_n$ to a given 
spin structure.

The definition of the Dirac operator $\mathrm{D}$ rests on the canonical 
connection on $Q$ that is given by the lift of the Levi-Civita connection 
on $FM$. Denoting the space of smooth sections of $S$ by $C^\infty (S)$, 
this connection induces a covariant derivative 
$\nabla^S:C^\infty (S)\rightarrow C^\infty(S\otimes T^*M)$. One then
defines $\mathrm{D}=\mu\circ\iota \circ\nabla^S$, where $\iota$ denotes 
the canonical isomorphism between $T^*M$ and $TM$ induced by the metric 
$\mathbf{g}$, and $\mu$ is the Clifford multiplication of a vector field 
and a smooth section in $S$.

The space $C_0^\infty (S)$ of smooth, compactly supported sections in $S$ 
can be turned into a pre-Hilbert space by introducing the inner product 
\begin{equation*}
 \langle\Psi,\Phi\rangle_{\mathrm{L}^2} := 
 \int_M\langle \Psi(m),\Phi(m)\rangle_{\mathbb{C}^{2^{n/2}}}\df \mu(m) \ ,
\end{equation*}
where $\df \mu(m)$ denotes the volume form on $M$. The resulting Hilbert 
space will be denoted by $\mathrm{L}^2(S)$. The Dirac operator $\mathrm{D}$
is elliptic and essentially self-adjoint on $C_0^\infty (S)$, thus its
spectrum is real. Moreover, on a compact manifold $M$ the spectrum is
discrete. 

In the sequel we will focus on compact surfaces of constant negative 
curvature. Their universal covering space is the upper half-plane 
$\mathbb{H}^2:=\{(x,y)\in \mathbb{R}^2 |y>0\}$ endowed with the Poincar\'e
metric $\mathbf{g}=y^{-2}(\df x\otimes \df x +\df y\otimes \df y)$ of
constant negative Gaussian curvature $K=-1$. The group of 
orientation-preserving isometries 
$\mathrm{PSL}(2,\mathbb{R})=\mathrm{SL}(2,\mathbb{R})/\{\pm\mathrm{Id}\}$ 
acts on $\mathbb{H}^2$ via fractional linear transformations,
\begin{equation*}
 z \mapsto \gamma z = \frac{az+b}{cz+d} \ , 
\end{equation*}
where $\gamma= (\begin{smallmatrix} a&b\\c&d \end{smallmatrix})\in
\mathrm{SL}(2,\mathbb{R})$ is a representative for an element in 
$\mathrm{PSL}(2,\mathbb{R})$ and $z=x+\imagunit y$ is viewed as a point
in $\mathbb{H}^2$. Any compact hyperbolic surface can now be represented as
$\Gamma\backslash\mathbb{H}^2$, where 
$\Gamma\subset \mathrm{PSL}(2,\mathbb{R})$ is a discrete, strictly
hyperbolic subgroup (a cocompact Fuchsian group of the first kind).

On the simply connected upper half-plane the relevant bundles are trivial,
i.e., $F\mathbb{H}^2\simeq \mathbb{H}^2\times\mathrm{SO}(n)$ and 
$Q\simeq\mathbb{H}^2\times\mathrm{Spin}(n)$. This allows an immediate 
construction of the Dirac operator, which is explicitly given as the 
matrix-valued operator
\begin{equation*}
 \mathrm{D}=\imagunit \begin{pmatrix} 
 0 & \imagunit y\frac{\partial}{\partial x}+y\frac{\partial}{\partial y}
 -\frac{1}{2} \\
 -\imagunit y\frac{\partial}{\partial x}+y\frac{\partial}{\partial y}-
 \frac{1}{2} & 0 \end{pmatrix} \ ,
\end{equation*}
see, e.g., also \cite{Pnueli:1994}. 
The operators appearing in the off-diagonals have been introduced by 
Maa\ss\ and have been further investigated by Roelcke
(\cite{Maass:1953,Roelcke:1966,Roelcke:1967}). In fact, they considered 
the operators
\begin{eqnarray}
\mathrm{K}_{k} 
  &:=& \imagunit y\frac{\partial}{\partial{x}}+y\frac{\partial}{\partial{y}}
       +\frac{k}{2} \ , \nonumber \\
\mathrm{\Lambda}_{k} 
  &:=& \imagunit y\frac{\partial}{\partial{x}}-y\frac{\partial}{\partial{y}}
       +\frac{k}{2} \ . \nonumber
\end{eqnarray}
It is therefore natural to introduce the weighted Dirac operators
\begin{equation}
\label{eq:DiracMaass}
 \mathrm{D}_k := \imagunit \begin{pmatrix}
  0 & \mathrm{K}_{k-2} \\ -\Lambda_k & 0 \end{pmatrix}, 
\end{equation}
which reduce to the Dirac operator for $k=1$. In physical terms the 
additional parameter $k$, which from now on will be called a weight, 
corresponds to a constant magnetic field on the surface. In normalised 
units the field strength is given by 
\begin{equation}
\label{magnetic}
 B=\frac{k-1}{2e} \ .
\end{equation}
On a compact surface the Dirac quantisation condition for the magnetic flux 
implies that $k$ must be an integer. Therefore, from now on we will only
consider $k\in\mathbb{Z}$.

The weighted Dirac operator $\mathrm{D}_k$ is closely related to the 
weighted Maa\ss -Laplacian,
\begin{equation*}
 \Delta_k = -\mathrm{K}_{k-2}\Lambda_k -\tfrac{k}{2}(1-\tfrac{k}{2})
          = y^2(\tfrac{\partial^2}{\partial x^2}  +  
            \tfrac{\partial^2}{\partial y^2})-\imagunit k y 
            \tfrac{\partial}{\partial x} \ ,
\end{equation*}
since
\begin{equation}
\label{eq:square}
 \mathrm{D}^{2}_{k}= 
 \begin{pmatrix}
 -\Delta_{k}-\frac{k}{2}\left(1-\frac{k}{2}\right) & 0 \\
 0 & -\Delta_{k-2}-\frac{k}{2}\left(1-\frac{k}{2}\right)
 \end{pmatrix} \ .
\end{equation}
We remark that both the weighted Laplacians and 
the weighted Dirac operators are elliptic operators as their principal 
symbols do not depend on $k$.

As yet we have neither specified spaces of sections in spinor bundles
$S$, nor have we introduced Dirac operators on compact surfaces. It turns
out that both purposes can be conveniently achieved by identifying
suitable automorphic forms. We devote the following section to this
task.
\section{Automorphic forms}
\label{section:auto}
All relevant bundles over a compact surface $\Gamma\backslash\mathbb{H}^2$
correspond to trivial bundles over $\mathbb{H}^2$ 
(see \cite{Gunning:1956,Gunning:1976}). More specifically, 
given the canonical projection 
$p:\mathbb{H}^2\rightarrow \Gamma\backslash\mathbb{H}^2$ and the vector bundle
$\zeta=(E,\Gamma\backslash\mathbb{H}^2,\pi,V)$ over the compact surface, 
its pullback $p^*(\zeta)$ is trivial, i.e., the total space is isomorphic to 
$\mathbb{H}^2\times V$. Now any covering translation $\gamma\in \Gamma$ 
extends to a bundle homomorphism $\tilde{\gamma}$ on $p^*(\zeta)$, and this 
can be written as 
\begin{equation}
\label{eq:factaut}
 \tilde{\gamma}(z,v)=(\gamma z,\sigma(z,\gamma)v)\ ,
\end{equation}
where $\sigma : \mathbb{H}^2\times \Gamma \rightarrow \mathrm{GL}(V)$.

For our purposes it will be useful to consider the subgroup 
$\bar{\Gamma}\subset\mathrm{SL}(2,\mathbb{R})$ that corresponds to the
Fuchsian group $\Gamma\subset\psl$ via 
$\Gamma=\bar{\Gamma}/\{\pm\mathrm{Id}\}$.
Instead of $\sigma$ one hence considers a function 
$j:\mathbb{H}^2\times\bar{\Gamma}\rightarrow \mathrm{GL}(V)$, called factor
of automorphy. To define the action of $\gamma \in \Gamma$ unambiguously 
one therefore requires a suitable character $\chi$ of $\bar{\Gamma}$ 
(multiplier system) depending on the value of $j(z,-\mathrm{Id})$.
Based on these observations sections in $\zeta$ can be obtained from
global sections in $p^*(\zeta)$: any vector-valued function 
$\psi: \mathbb{H}^2 \to V$ with 
$\psi(\gamma z)=\chi(\gamma)\,j(z,\gamma)\,\psi(z)$, called automorphic
form, yields a section in $\zeta$, and vice versa.
\begin{defn}
Let $\Gamma\subset\psl$ be a strictly hyperbolic Fuchsian group, then a 
multiplier system of weight $k$ is a unitary character $\chi$ of 
$\bar{\Gamma}$ with $\chi(-\mathrm{Id}) = (-1)^k$. A factor of automorphy 
for the Maa\ss -Laplacian of weight $k$ is defined by
$j_{\gamma}(z,k):=\tfrac{(cz+d)^{k/2}}{(c\bar{z}+d)^{k/2}}$ when
$\gamma=\left(\begin{smallmatrix}a&b\\c&d\end{smallmatrix}\right)$. Then 
$\psi:\mathbb{H}^2\rightarrow \mathbb{C}$ is an automorphic form for the
Maa\ss -Laplacian of weight $k$, if
\begin{equation*}
 \psi(\gamma z) = \chi(\gamma)\,j_{\gamma}(z,k)\,\psi(z) \ ,
 \quad \forall \gamma\in  \bar{\Gamma} \ .
\end{equation*}
The space of all such automorphic forms is denoted by 
$^L\mathfrak{F}(\Gamma,k,\chi)$. Similarly, $^L\mathfrak{C}^l(\Gamma,k,\chi)$ 
and $^L\negthinspace \mathfrak{L}^2(\Gamma,k,\chi)$ are the spaces of 
differentiable and square-integrable automorphic forms, respectively.
In the latter case, integrability is meant with respect to the measure 
$\df \mu(z)=\tfrac{\df x\df y}{y^2}$ over a fundamental domain 
$F\subset\mathbb{H}^2$ of $\Gamma$.
\end{defn}
It is well-known that $\Delta_k$ acting on a (twice differentiable) 
automorphic form of weight $k$ yields again an automorphic form of weight 
$k$ (see \cite{Roelcke:1966,Roelcke:1967,Hejhal:1976}). We denote the space 
of all eigenforms of $-\Delta_k$ with eigenvalue $\lambda$ by 
$^L\negthinspace\mathfrak{F}_{\lambda}(\Gamma,k,\chi)$. In case $k\geq 2$
there exist certain special eigenvalues that are known explicitly,
\begin{equation}
\label{eq:specialev}
 \lambda^{(k)}_j = \frac{k-2j}{2}\left(1-\frac{k-2j}{2}\right)\ ,
 \quad j=0,1,2,\dots, \left[\frac{k-1}{2}\right]\ ,
\end{equation}
where $[x]$ is the integer part of $x\in\mathbb{R}$. For multiplier systems
$\chi\not\equiv 1$ the eigenvalue (\ref{eq:specialev}) has a multiplicity
$(g-1)(k-2j-1)$, see \cite{Hejhal:1976}; here $g\geq 2$ is the genus of the 
closed surface $\Gamma\backslash\mathbb{H}^2$.

The relations (\ref{eq:DiracMaass}) and (\ref{eq:square}) now suggest the
following definition of automorphic forms for the weighted Dirac operator.
\begin{defn}
Let
\begin{equation*}
 J_{\gamma}(z,k) := 
 \begin{pmatrix} j_{\gamma}(z,k) & 0 \\ 0 & j_{\gamma}(z,k-2)\end{pmatrix} \ .
\end{equation*}
Then we define the space $\mathfrak{F}(\Gamma,k,\chi)$ of automorphic forms
for the Dirac operator with weight $k$ to consist of the functions
$\Psi:\mathbb{H}^2\rightarrow \Delta_2 = \mathbb{C}^2$ that transform as
\begin{equation}
\label{eq:Diracauto}
 \Psi(\gamma z)=\chi(\gamma)\,J_{\gamma}(z,k)\,\Psi(z) \ ,
 \quad \forall \gamma\in\bar{\Gamma} \ .
\end{equation}
The spaces $\mathfrak{C}^l(\Gamma,k,\chi)$, $\mathfrak{L}^2(\Gamma,k,\chi)$ 
are defined analogously, and $\mathfrak{F}_{\rho}(\Gamma,k,\chi)$ denotes 
the eigenspace of $-\mathrm{D}_k$ corresponding to the eigenvalue $\rho$.
\end{defn}
In order to demonstrate that the weighted Dirac operator $\mathrm{D}_k$ 
maps a differentiable automorphic form of the type (\ref{eq:Diracauto}) 
to a form of the same type, one has to show that under a covering
translation $\gamma$ the operator $\mathrm{D}_k$ behaves as
\begin{equation}
\label{eq:traf}
 \mathrm{D}_k\mapsto 
 J_{\gamma}(z,k)\,\mathrm{D}_k \,J^{-1}_{\gamma}(z,k)\ .
\end{equation}
This can be done in a straightforward calculation. Hence, the automorphic
forms (\ref{eq:Diracauto}) represent sections in spinor bundles over the
surface $\Gamma\backslash\mathbb{H}^2$.

We also note that $\mathrm{D}_k$ is essentially self-adjoint on 
$\mathfrak{C}^1(\Gamma,k,\chi)$. The proof is practically identical to
the case of the Dirac operator ($k=1$), which is well-known (see, e.g.,
\cite{Friedrich:1997} where also $\mathrm{Spin}^{\mathbb{C}}$-connections
are considered). Thus we conclude that the spectrum of $\mathrm{D}_k$ is 
real and discrete.

We are now able to establish a connection between the eigenforms
of the Laplacian and of the Dirac operator. For this we recall that
the spectrum of $-\Delta_k$ is bounded from below by 
$\tfrac{k}{2}(1-\tfrac{k}{2})$ \cite{Roelcke:1966}.
\begin{satz}
\label{satz:auto}
Let $\Psi=\left(\begin{smallmatrix} \psi_1\\ \psi_2
\end{smallmatrix}\right)$ be an eigenform of $-\mathrm{D}_k$ with eigenvalue
$\rho$, and let $\lambda=\rho^2+\tfrac{k}{2}(1-\tfrac{k}{2})$.
Then $\psi_1 \in \,^L\negthinspace\mathfrak{F}_{\lambda}(\Gamma,k,\chi)$.
On the other hand, if 
$\psi\in\, ^L\negthinspace\mathfrak{F}_{\lambda}(\Gamma,k,\chi)$,
then 
$\Psi=\left(\begin{smallmatrix} \rho \psi \\ \imagunit \Lambda_k \psi
\end{smallmatrix}\right)\in  \mathfrak{F}_{\rho}(\Gamma,k,\chi)$.
Moreover, if eigenforms of the Dirac operator ($\rho\neq0$) are linearly 
independent, the same holds for the corresponding eigenforms of the 
Maa\ss -Laplacian, and vice versa.
\end{satz}
\begin{proof}
As both operators are elliptic all eigenforms are smooth. The relation 
between $\rho$ and $\lambda$ can be obtained by using eq.~(\ref{eq:square}).
To establish the desired transformation properties one has to write down
the conditions on the components. The rest then follows from 
\cite{Roelcke:1966}. The linear independence can be obtained from some 
straight-forward manipulations using
\begin{equation*}
 -\imagunit \Lambda_k \psi_1+\rho\psi_2=0 \ ,
\end{equation*}
which can be solved for $\psi_2$ if $\rho\neq0$.
\end{proof}
\section{Symmetries}
\label{section:sym}
Apart from the fact that spectra of weighted Dirac operators on compact 
surfaces are real and discrete, further spectral properties can be concluded 
from symmetry considerations.
\begin{lemma}
\label{lem:chiral}
The weighted Dirac operator possesses a chiral symmetry, i.e., if
$\Psi=\left(\begin{smallmatrix} \psi_1 \\ \psi_2\end{smallmatrix}\right)
 \in\mathfrak{F}_{\rho}(\Gamma,k,\chi)$, then
$\left(\begin{smallmatrix} \psi_1 \\ -\psi_2\end{smallmatrix}\right) 
\in\mathfrak{F}_{-\rho}(\Gamma,k,\chi)$. Hence the spectrum of $\mathrm{D}_k$
is symmetric about zero. 
\end{lemma} 
\begin{proof}
The proof amounts to a simple calculation which can be performed conveniently 
by using the relations in \cite{Roelcke:1966} once again.
\end{proof}
The symmetry statement extends to zero modes in that the Atiyah-Singer 
index theorem (see \cite{Hitchin:1974,Atiyah:1968}) implies that their 
number is even. From now on we will denote this number by $2N$.

A further symmetry is concerned with quantum mechanical time-reversal.
For the Dirac operators considered here this is implemented through 
the anti-linear operator $\mathrm{T}:=\imagunit \sigma_2 \mathrm{C}$ acting 
on automorphic forms of arbitrary weight, where $\mathrm{C}$ denotes complex 
conjugation and $\sigma_2 = \left(\begin{smallmatrix} 0& -\imagunit \\ 
\imagunit &0 \end{smallmatrix}\right)$ is one of the Pauli matrices. 
Hence $\mathrm{T}^2 = -\mathrm{Id}$. Moreover, on the Hilbert space 
$\mathfrak{L}^2(\Gamma,k,\chi)$ this operator is anti-unitary.
\begin{lemma}
\label{lem:timereversal}
Let $\Psi\in \mathfrak{F}_{\rho}(\Gamma,k,\chi)$, then 
$\mathrm{T}\Psi\in \mathfrak{F}_{\rho}(\Gamma,2-k,\bar{\chi})$. 
In particular, if $k=1$ and the multiplier system $\chi$ is real-valued
every eigenvalue of $\mathrm{D}$ has a multiplicity of (at least) two.
\end{lemma}
\begin{proof}
The first part of this lemma can be checked by straight-forward calculations 
that we omit here. For the second part one merely has to use the fact 
that $\mathrm{T}$ is anti-unitary with $\mathrm{T}^2 = -\mathrm{Id}$. 
This immediately yields that the eigenforms $\Psi$ and $\mathrm{T}\Psi$ 
with eigenvalue $\rho$ are linearly independent. 
\end{proof}
At this point we add a few remarks:
\begin{itemize}
\item[(i)]
Above we have always assumed that $\Gamma$ contains, apart from unity, only 
hyperbolic elements; hence there are no elements of finite order. Thus one 
can construct a real-valued multiplier system by simply assigning 
$\pm 1$ to the generators of $\Gamma$ and extending this to $\bar{\Gamma}$. 
If one also allows for elliptic elements this clearly will not work.
\item[(ii)]
Lemma~\ref{lem:timereversal} says that $\mathrm{D}_k$ and $\mathrm{D}_{2-k}$
possess identical non-vanishing eigenvalues (including multiplicities).
According to (\ref{magnetic}) the weights $k$ and $2-k$, up to sign 
correspond to the same magnetic field strengths. We will therefore restrict 
subsequent discussions to the case $k\geq 1$.
\item[(iii)]
In general, quantum systems with half-integer spin share the same behaviour
under time-reversal as in the present case. For the respective quantum 
Hamiltonians the multiplicity two of their eigenvalues due to time-reversal 
symmetry is known as Kramers' degeneracy \cite{Kramers:1930}. Moreover, 
magnetic fields usually break time-reversal invariance. This is the reason 
for the special role played by the weight $k=1$.
\item[(iv)]
In his tenfold-way scheme Zirnbauer classified quantum systems according
to their behaviour under basic symmetry operations like time-reversal, 
rotations, and chiral transformations \cite{Zirnbauer:1996}. The resulting 
symmetry classes are unambiguously linked with Cartan's ten classes of 
symmetric spaces. In this context the present case, with time-reversal 
symmetry, $\mathrm{T}^2 = -\mathrm{Id}$, and chiral symmetry is identified 
as the type $C\mathrm{II}$ and is related to the symmetric space 
$\mathrm{Sp}(p+q)/\mathrm{Sp}(p)\times\mathrm{Sp}(q)$ (of compact type). 
\item[(v)]
Following the conjecture of Bohigas, Giannoni, and Schmit \cite{Bohigas:1984}
one expects that, generically, correlations among the eigenvalues of geometric 
operators on manifolds of negative curvature can be described by random matrix
theory. (See, however, exceptions when the Fuchsian group is arithmetic
\cite{Bogomolny:1992,Bolte:1992}.) 
The relevant random matrix ensemble for the present case is
the chiral Gaussian symplectic ensemble (chGSE). 
 \end{itemize}
From supersymmetry considerations (see, e.g., \cite{Thaller:1992}) 
it is known that eigenvalues of the square of a Dirac operator have
even multiplicities. We stress that this observation is unrelated to 
Lemma~\ref{lem:timereversal}. Rather, this degeneracy stems from chiral
symmetry (Lemma~\ref{lem:chiral}) combined with the squaring.

However, the degeneracy due to time reversal somewhat unexpectedly extends 
to the spectrum of $\mathrm{D}_k$ for odd weight. To this end we recall that 
the Maa\ss\ operator $\mathrm{K}_k$ raises the weight by two, without changing 
the eigenvalue of an eigenform of the appropriate Maa\ss -Laplacian 
\cite{Roelcke:1966,Roelcke:1967}. It hence maps
$^L\negthinspace\mathfrak{F}_{\lambda}(\Gamma,k,\chi)$ to 
$^L\negthinspace\mathfrak{F}_{\lambda}(\Gamma,k+2,\chi)$. The operator
$\Lambda_{k+2}$ is its formal adjoint from 
$^L\negthinspace \mathfrak{L}^2(\Gamma,k+2,\chi)$ to
$^L\negthinspace \mathfrak{L}^2(\Gamma,k,\chi)$. As a consequence,
up to the special eigenvalues (\ref{eq:specialev}) the spectrum of
$-\Delta_k$ depends only on $k$ mod $2$. 
\begin{lemma}
\label{lemma:Maassmult}
Let $\Gamma$ be a strictly hyperbolic group and $\chi$ a real-valued 
multiplier system. If the weight $k$ is odd the eigenvalues of the 
Maa\ss -Laplacian $\Delta_k$ have a multiplicity of (at least) two.
\end{lemma}
\begin{proof}
For odd $k=2m+1$, the multiplicities of the special eigenvalues 
(\ref{eq:specialev}) are known to be $(g-1)2(m-j)$, see \cite{Hejhal:1976}
and below equation (\ref{eq:specialev}). In all other cases the statement
follows immediately from Proposition~\ref{satz:auto} and 
Lemma~\ref{lem:timereversal}. 
\end{proof}
Apart from the special eigenvalues, a twofold degeneracy of the eigenvalues
can also be obtained in a constructive way. To this end, one first
lowers the weight from $k=2m+1$, $m\in\mathbb{N}_0$, to one by successive 
applications of operators $\Lambda_l$, then applies complex conjugation 
(i.e., time reversal for spin zero), and finally raises the weight back to 
$k=2m+1$. Altogether, one thus applies the anti-linear operator
\begin{equation}
\label{eq:sym}
 ^L\negthinspace S_{2m+1}:=
 \mathrm{C}\Lambda_{-2m+1}\Lambda_{-2m+3}\dots\Lambda_{2m+1} \ .
\end{equation}
Now some straight-forward calculations show that this way an eigenform
$\psi$ is mapped to a linearly independent eigenform 
$\phi= \, ^L\negthinspace S_{2m+1}\psi$ of the same weight and with
the same eigenvalue. In particular, $\phi$ may only vanish when its 
eigenvalue is of the special type (\ref{eq:specialev}).

We stress that this argument fails if $k$ is even. In the case $k=0$ this is 
easy to see: if $\psi\in\,^L\negthinspace\mathfrak{F}_{\rho}(\Gamma,0,\chi)$, 
then $\mathrm{C}\psi\in \,^L\negthinspace\mathfrak{F}_{\rho}(\Gamma,0,\chi)$ 
as long as $\chi$ is real. But it is well-known that for $k=0$ the eigenforms 
can be chosen to be real-valued, so $\psi$ and $\mathrm{C}\psi$ 
are not linearly independent.

Of course, according to Proposition~\ref{satz:auto} the degeneracy in the 
spectra of Maa\ss-Laplacians implies a corresponding degeneracy in the 
spectra of the Dirac operators $\mathrm{D}_k$. Except for the eigenvalues
that derive from the special eigenvalues (\ref{eq:specialev}) of the
associated Maa\ss-Laplacian, one can again devise a constructive approach 
which reveals more clearly that the degeneracy is a consequence of 
a `generalised time reversal symmetry'. To this end we introduce the operator
\begin{equation*}
 \mathrm{A}_k^{\dagger}:= \begin{pmatrix} 
 \rho'  \mathrm{K}_k & 0\\ \imagunit k &\rho \mathrm{K}_{k-2}
 \end{pmatrix} \ .
\end{equation*}
This maps $\mathfrak{F}_{\rho}(\Gamma,k,\chi)$ into 
$\mathfrak{F}_{\rho'}(\Gamma,k+2,\chi)$, where $\rho'$ is determined by 
$\rho^2+k=\rho'^2$ and $\mathrm{sgn}(\rho')=\mathrm{sgn}(\rho)$.
From \cite[Satz 5.3]{Roelcke:1966} one can deduce that 
$\mathrm{A}_k^{\dagger}$ indeed is an isomorphism between
$\mathfrak{F}_{\rho}(\Gamma,k,\chi)$ and 
$\mathfrak{F}_{\rho'}(\Gamma,k+2,\chi)$ if $\rho\neq 0$. Its formal 
adjoint reads 
\begin{equation*}
 \mathrm{A}_{k+2}:=- \begin{pmatrix}
 \rho'  \mathrm{\Lambda}_{k+2} & \imagunit k\\0 &\rho \mathrm{\Lambda}_{k}
 \end{pmatrix} \ .
\end{equation*}
The analogue to the operator (\ref{eq:sym}) now is
\begin{equation*}
 S_{2m+1} := \mathrm{B}_{2m-1}^{\dagger}\dots\mathrm{B}_1^{\dagger}
 \mathrm{T}\mathrm{C}_3\dots\mathrm{C}_{2m+1} \ ,
\end{equation*}
where $\mathrm{B}^{\dagger}_k:=\tfrac{1}{\rho}\mathrm{A}^{\dagger}_k$
and $\mathrm{C}_k=\tfrac{1}{\rho'}\mathrm{A}_k$. A direct calculation shows 
that $\mathrm{S}_{2m+1}$ does not depend on $\rho$. 
\begin{satz}
Let $k=2m+1$ and assume that the multiplier system $\chi$ is real-valued.
Then the eigenvalues of the weighted Dirac operator $\mathrm{D}_{2m+1}$
occur with multiplicities of at least two. Moreover,
\begin{equation*}
 \mathrm{S}_{2m+1}=\mathrm{T}\, \begin{pmatrix}
 \Lambda_{-2m+3}\Lambda_{-2m+5}\dots\Lambda_{2m+1}& 0 \\
 0 & \Lambda_{-2m+1}\Lambda_{-2m+3}\dots\Lambda_{2m-1} 
 \end{pmatrix} 
\end{equation*}
is identical on all spaces $\mathfrak{F}_{\rho}(\Gamma,2m+1,\chi)$ of 
eigenforms, and can therefore be extended to all of 
$\mathfrak{C}^\infty(\Gamma,2m+1,\chi)$. Then
$\Psi\in \mathfrak{F}_{\rho}(\Gamma,2m+1,\chi)$ implies 
$\Phi:=\mathrm{S}_{2m+1}\Psi \in\mathfrak{F}_{\rho}(\Gamma,2m+1,\chi)$
and $\langle\Phi,\Psi\rangle_{\mathrm{L}^2}=0$. For $m>0$ the form 
$\Phi$ vanishes identically iff $\rho$ is related to one of the special 
eigenvalues (\ref{eq:specialev}) via 
$\lambda^{(k)}_j=\rho^2+\tfrac{k}{2}(1-\tfrac{k}{2})$.
\end{satz}
\begin{proof}
The statement about the multiplicities of eigenvalues is implied by
Proposition~\ref{satz:auto} and Lemma~\ref{lemma:Maassmult}.
The conclusions $\Phi\in \mathfrak{F}_{\rho}(\Gamma,k,\chi)$ and 
$\langle\Phi,\Psi\rangle_{\mathrm{L}^2}=0$ follow directly from the 
preceding considerations. One must only take care of the possibility that
$\Phi$ may vanish. But evaluating $\|\Phi\|_{\mathrm{L}^2}$ shows that 
this norm is zero, iff $\lambda=\rho^2+\tfrac{k}{2}(1-\tfrac{k}{2})$
is of the form (\ref{eq:specialev}).
\end{proof}
\section{Point-pair invariants}
\label{section:point-pair}
From the case of the Laplacian it is well known that setting up a Selberg
trace formula amounts to calculating traces of Hilbert-Schmidt operators 
$\mathrm{L}$ that commute with the Laplacian 
\cite{Selberg:1956,Hejhal:1976,Hejhal:1983}. The kernels of such operators
can be obtained from suitable Poincar\'e series over so-called point-pair 
invariants, whose construction will be briefly outlined in this section.

We first introduce two matrix-valued functions on the upper half-plane,
which help to study transformation properties under fractional linear
transformations,
\begin{eqnarray}
A(z',z) &:=& \begin{pmatrix}
   \frac{(z-\bar{z}')^{\frac{1}{2}}}{(z'-z)^{\frac{1}{2}}} & 0 \\
   0 & \frac{(z'-\bar{z})^{\frac{1}{2}}}{(\bar{z}-\bar{z}')^{\frac{1}{2}}} 
   \end{pmatrix}\nonumber \ ,\\
B(z',z) &:=& \begin{pmatrix}
   \frac{(z'-z)^{\frac{1}{2}}}{(z'-\bar{z})^{\frac{1}{2}}} & 0 \\
   0 &  \frac{(\bar{z}-\bar{z}')^{\frac{1}{2}}}{(z-\bar{z}')^{\frac{1}{2}}} 
   \end{pmatrix} \nonumber \ .
\end{eqnarray}
\begin{defn}
\label{defn:ppair}
Let $\Phi=\left(\begin{smallmatrix} \Phi_1 & \Phi_2 \\ 
\Phi_3 & \Phi_4\end{smallmatrix}\right):[1,\infty)\rightarrow 
\mathbb{R}^{2\times 2}$ be continuous. Assume moreover that $\Phi_2=\Phi_3$ 
and that there exists a constant $C_{\epsilon}$ such that each component 
satisfies
\begin{equation}
\label{eq:Phiprop}
 |\Phi_i(\sigma)|\leq C_{\epsilon}\sigma^{-1-\epsilon}
\end{equation}
for some $\epsilon>0$. Then we call 
\begin{equation*}
  K(z',z)=
  -\imagunit A(z',z)\Phi\left(\tfrac{1}{2}(\cosh(d(z',z)+1)\right)B(z',z)
  \ ,
\end{equation*}
where $d(z',z)$ denotes the hyperbolic distance of two points 
$z,z'\in\mathbb{H}^2$, a point-pair invariant.
\end{defn}
Applying any transformation $\gamma\in\mathrm{SL}(2,\mathbb{R})$ yields
\begin{equation*}
 K(\gamma z',\gamma z)=J_{\gamma}(z',1)\,K(z',z)\,J_{\gamma}^{-1}(z,1) \ .
\end{equation*}
The restrictions on the components of $\Phi$ assure that $K$ is hermitian, 
i.e., $K(z',z)=K^{\dagger}(z,z')$. In order to proceed we need a technical 
Lemma, which can be found in \cite{Hejhal:1976}.
\begin{lemma}
\label{lemma:tech}
Let $f:[1,\infty)\rightarrow \mathbb{C}$ and $\epsilon>0$. Moreover, assume
that
\begin{equation*}
 f(\sigma)\leq C_{\epsilon} \sigma^{-1-\epsilon}
\end{equation*}
holds for all $\sigma$ and a constant $C_{\epsilon}$. Then the sum
\begin{equation*}
 \sum_{\gamma\in\bar{\Gamma}}
 f\left(\tfrac{1}{2}(\cosh(d(z',\gamma z))+1)\right)
\end{equation*}
converges absolutely and uniformly for all $z',z\in\mathbb{H}^2$.
\end{lemma}
Starting from the point-pair invariant we can now construct automorphic 
kernels via Poincar\'e series. Due to our restrictions on $\Phi$,
\begin{equation}
\label{eq:automorphic}
 K_{\Gamma}(z',z)=\frac{1}{2}\sum_{\gamma\in\bar{\Gamma}}
 K(z',\gamma z)\,\chi(\gamma)\,J_{\gamma}(z,1)
\end{equation}
is well-defined. Some standard manipulations show that the behaviour of
$K_{\Gamma}$ under transformations is given by
\begin{equation}
\label{eq:transpp}
 K_{\Gamma}(\gamma_1z',\gamma_2z)=\chi(\gamma_1)\,J_{\gamma_1}
 (z',1)\,K_{\Gamma}(z',z)\,J_{\gamma_2}^{-1}(z,1)\,\chi^{-1}(\gamma_2) \ .
\end{equation}
This allows us to define automorphic kernels.
\begin{defn}
Let $F$ be a fundamental domain for the Fuchsian group $\Gamma$. We say that
$K_{\Gamma}:\mathbb{H}^2\times \mathbb{H}^2\rightarrow \mathbb{C}^{2\times 2}$
is an element of $\mathfrak{L}^2(\Gamma\backslash\mathbb{H}^2,
\Gamma\backslash\mathbb{H}^2,\chi)$, if (\ref{eq:transpp}) holds for all 
$\gamma\in \bar{\Gamma}$ and 
\begin{equation*}
 \|K_{\Gamma}\|^2_{\mathrm{L}^2}:=
 \int_F\limits \int_F\limits \mathrm{tr}\left(K^{\dagger}_{\Gamma}(z',z) 
 K_{\Gamma}(z',z)\right) \df \mu(z)\df \mu(z') <\infty \ .
\end{equation*}
\end{defn}
It is well-known that such an automorphic kernel defines a Hilbert-Schmidt 
operator $\mathrm{L}:\mathfrak{L}^2(\Gamma,1,\chi)\rightarrow 
\mathfrak{L}^2(\Gamma,1,\chi)$ via
\begin{equation*}
[\mathrm{L}\Psi](z'):=\int_{F}\limits K_{\Gamma}(z',z)\Psi(z)\df \mu(z) \ .
\end{equation*}
If $K_{\Gamma}$ is hermitian there exists a basis $\{\Psi_n\}$ of 
orthonormal eigenforms of $\mathrm{L}$ in $\mathfrak{L}^2(\Gamma,1,\chi)$
such that $K_{\Gamma}(z',z)=\sum_n\limits a_n\Psi_n(z')\Psi^{\dagger}_n(z)$.
Furthermore, if $K_{\Gamma}$ was constructed from a point-pair invariant 
via (\ref{eq:automorphic}), then one can easily check that the operator
$\mathrm{L}$ has a finite trace that is given by $\sum_n\limits a_n$.
In the following section we will construct point-pair invariants from
a Green's function and then calculate such traces.
\section{The trace formula}
\label{section:trace}
Before we can proceed to introduce point-pair invariants we have to identify 
Green's functions for Dirac operators on surfaces 
$\Gamma\backslash\mathbb{H}^2$. For this, and the following, we restrict our 
attention to the Dirac operator itself, i.e., to the weight $k=1$. The
starting point will be Green's function for $\mathrm{D}$ on the hyperbolic
plane, from which the corresponding Green's function on the compact surface
can be obtained in terms of a Poincar\'e series.

Since the resolvent of $\mathrm{D}$ is a bounded operator when
$\mathrm{Im}(\rho)<0$, we can make the ansatz
\begin{equation*}
 \left(\mathrm{D}+\rho \right)^{-1}\Psi(z')=
 \int_{\mathbb{H}^2} G(z',z;\rho)\Psi(z)\,\mbox{d}\mu(z) \ ,
\end{equation*}
with $G(\cdot,\cdot;\rho):\mathbb{H}^2\times\mathbb{H}^2\to
\mathbb{C}^{2\times 2}$. For the matrix entries of $G$ we use the notation
$G=\left(\begin{smallmatrix} G_1 & G_2\\ G_3 & G_4\end{smallmatrix}\right)$.
Then $G$ can be uniquely characterised as a solution of the 
matrix-differential equation
\begin{equation}
\label{eq:green}
 \left(\mathrm{D}+\rho \right)G(z',z;\rho)=0\ ,\quad\text{for}\ z\neq z',
\end{equation}
with a specified behaviour in a neighbourhood of $z=z'$. That is, for the 
diagonal matrix entries
\begin{equation*}
 \lim_{d(z',z)\rightarrow 0}\limits \left(G_i(z',z;\rho)-\frac{\rho}{4\pi}
 \log(d(z',z))\right)<\infty, \quad i=1,4\ ,
\end{equation*}
is required, whereas the non-diagonal entries are regular in $z=z'$. In
addition, $G_i(z',z;\rho)$ must approach zero as $d(z',z)\to\infty$.

Furthermore, the transformation property (\ref{eq:traf}) implies the
corresponding behaviour
\begin{equation*}
 G(z',z;\rho)=J_{\gamma}^{-1}(z',1)\,G(\gamma z',\gamma z;\rho)\, 
 J_{\gamma}(z,1) 
\end{equation*}
of the Green's function under an isometry 
$\gamma\in\mathrm{PSL}(2,\mathbb{R})$.

In order to solve for this Green's function we closely follow 
\cite{Roelcke:1966,Roelcke:1967}, where the corresponding problem for the
Maa\ss -Laplacians is treated. We begin with introducing 
\begin{equation*}
\check{G}(z',z;\rho):=A^{-1}(z',z)G(z',z;\rho)B^{-1}(z',z) \ ,
\end{equation*}
which is invariant under $\mathrm{PSL}(2,\mathbb{R})$, and note that
$\check{\mathrm{D}} := A^{-1}(z,\imagunit)\,\mathrm{D}\,A(z,\imagunit)$
is invariant under the stability group $\mathrm{PSO}(2,\mathbb{R})$ of 
$z=\imagunit$. Therefore, the differential equation for $\check{G}$ 
corresponding to (\ref{eq:green}) is transformed into polar coordinates 
$(\sigma,\phi)$ for $z\in\mathbb{H}^2$, where 
$\sigma=\tfrac{1}{2}(\cosh(d(z,z'))+1)$. With 
$H(\sigma;\rho)=\check{G}(z',z;\rho)$, this leads to a system of ordinary 
linear differential equations for
$H=\left(\begin{smallmatrix} H_1 & H_2\\ H_3 & H_4\end{smallmatrix}\right)$,
\begin{multline}
\label{eq:greenh}
 \left(\begin{smallmatrix}
 \rho & \imagunit \left[\left(\sigma(\sigma -1)\right)^{\frac{1}{2}}
 \partiell{\sigma}+\frac{1}{2}\left(\frac{\sigma -1}{\sigma}
 \right)^{\frac{1}{2}}\right]\\
 \imagunit \left[\left(\sigma(\sigma-1)\right)^{\frac{1}{2}}
 \partiell{\sigma}+\frac{1}{2}\left(\frac{\sigma -1}{\sigma}
 \right)^{\frac{1}{2}}\right] & \rho 
 \end{smallmatrix}\right)H(\sigma;\rho) \\
 = -\imagunit \left(\begin{smallmatrix}
 \frac{1}{2}\left(\sigma(\sigma-1)\right)^{-\frac{1}{2}}H_3(\sigma;\rho) & 0\\ 
 0 & \frac{1}{2}\left(\sigma(\sigma-1)\right)^{-\frac{1}{2}}H_2(\sigma;\rho)
 \end{smallmatrix}\right) \ .
\end{multline}
The solution of (\ref{eq:greenh}) for $\sigma\rightarrow\infty$ is given by
\begin{eqnarray}
\label{eq:Hdiag}
 H_1(\sigma;\rho)
 &=& -\frac{\rho}{4\pi}\sigma^{-\frac{1}{2}-\imagunit \rho}
     \frac{\Gamma(\imagunit\rho)\Gamma(\imagunit\rho+1)}{\Gamma(2\imagunit 
     \rho+1)}F(\imagunit \rho,1+\imagunit \rho;1+2\imagunit\rho;
     \tfrac{1}{\sigma})\nonumber \\
 &=& -\frac{\rho}{4\pi} \sigma^{-\frac{1}{2}}\int_0^1\limits 
     t^{\imagunit \rho} (1-t)^{\imagunit \rho -1} 
     (\sigma-t)^{-\imagunit \rho} \df t \ , 
\end{eqnarray}
and
\begin{eqnarray}
\label{eq:Hnondiag}
 H_2(\sigma;\rho)
 &=& -\frac{\imagunit}{4\pi}\sigma^{-1-\imagunit \rho}(\sigma-1)^{\frac{1}{2}}
     \frac{\Gamma^2(\imagunit\rho+1)}{\Gamma(2\imagunit \rho+1)}
     F(1+\imagunit \rho,1+\imagunit \rho;1+2\imagunit\rho;\tfrac{1}{\sigma})
     \nonumber \\
 &=& \frac{\rho}{4\pi} (\sigma-1)^{\frac{1}{2}}\int_0^1\limits 
     t^{\imagunit \rho}(1-t)^{\imagunit \rho -1} 
     (\sigma-t)^{-\imagunit \rho-1} \df t \ , 
\end{eqnarray}
where $F(a,b;c;z)$ is a hypergeometric function (see, e.g., 
\cite{Erdelyi:1953}). Moreover, $H_3(\sigma;\rho) = H_2(\sigma;\rho)$ and
$H_4(\sigma;\rho) = H_1(\sigma;\rho)$. From this representation we infer 
an upper bound for the components of the Green's function,
\begin{equation}
\label{eq:bound}
 |G_i(z',z;\rho)\leq const.\ |\rho|\,
 \ez^{-(\frac{1}{2}-\mathrm{Im}(\rho))d(z',z)}\ , 
 \quad\text{if}\ d(z',z)\geq d_0>0\ .
\end{equation}
As the singularity of $G$ for $d(z',z)\rightarrow 0$ is integrable we obtain 
the following Lemma.
\begin{lemma}
\label{lemma:techver}
Let $f:\mathbb{H}^2\rightarrow \mathbb{C}^2$ be bounded and continuous, then
\begin{equation*}
 \int_{\mathbb{H}^2}\limits G(z',z;\rho) f(z)\df \mu(z)
\end{equation*}
converges absolutely and uniformly in $\mathrm{Re}(\rho)$ as long as 
$\mathrm{Im}(\rho)<-\frac{1}{2}$.
\end{lemma}
With Lemma~\ref{lemma:tech} and the estimate (\ref{eq:bound}), Green's 
function for the Dirac operator on a compact surface can now be obtained 
in terms of a Poincar\'e series (\ref{eq:automorphic}).
\begin{lemma}
Let $\mathrm{Im}(\rho)<-\tfrac{1}{2}$, then
\begin{equation*}
 G_{\Gamma}(z',z;\rho) :=
 \frac{1}{2}\sum_{\gamma \in\bar{\Gamma}}\limits G(z',\gamma z;\rho)
 \,\chi(\gamma)\,J_{\gamma}(z,1)
\end{equation*}
converges for $z\neq z'\mod {\Gamma}$, and is the Green's function for
$\mathrm{D}$ on $\Gamma\backslash\mathbb{H}^2$.
\end{lemma}
We are now in a position to construct point-pair invariants as in 
Definition~\ref{defn:ppair} through
\begin{equation*}
 \Phi(\sigma):=\frac{1}{\pi} \int_{-\infty}^{\infty}\limits 
 H(\sigma;\rho)\,h(\rho)\ \df \rho \ ,
\end{equation*}
where $h$ is a function as specified below.
\begin{defn}
\label{defn:h}
A function $h:\mathbb{C}\rightarrow \mathbb{C}$ that satisfies
\begin{itemize}
\item
$h$ is even, i.e. $h(\rho)=h(-\rho)$,
\item
$h$ is complex analytic in the strip $|\mathrm{Im}(\rho)|\leq \beta$ for some 
fixed $\beta\geq\tfrac{1}{2}+\epsilon$,
\item
there exists $\delta>0$, such that the bound
\begin{equation*}
 |h(\rho)| \leq const.\ (1+\mathrm{Re}(\rho))^{-2-\delta}
\end{equation*} 
holds uniformly for all $\rho$ in the above mentioned strip,
\end{itemize}
is called an admissible test function.
\end{defn}
We denote the Fourier transform of an admissible test function by
\begin{equation*}
 g(u):=\frac{1}{2\pi}\int_{-\infty}^{\infty}\limits h(\rho)\,
       \ez^{-\imagunit \rho u}\ \df \rho \ .
\end{equation*}
All that we have to check that this indeed leads to a point-pair invariant
is the condition~(\ref{eq:Phiprop}).
\begin{lemma}
\label{lemma:id}
Let $\sigma-1\geq \kappa>0$, then the components $K_i$ of $K$ are bounded
from above by
\begin{equation}
\label{eq:Kestimate}
 |K_i(z',z)|\leq C_{\kappa}\,\ez^{-(\frac{1}{2}+\beta) d(z',z)}\ , 
 \quad i=1,\dots,4 \ ,
\end{equation}
with some $C_{\kappa}>0$.
Furthermore, the limit $\lim_{d(z',z) \rightarrow 0} \limits K(z',z)$ is 
well-defined. More precisely, we have
\begin{equation*}
 \mathrm{tr}\,K(z,z)=\frac{1}{2\pi}\int_{-\infty}^{\infty}\limits
 \rho\,h(\rho)\,\coth(\pi\rho)\ \df \rho  \ .
\end{equation*}
\end{lemma}
\begin{proof}
To prove the first part one shifts the integral by $-\imagunit \beta$. 
Using the estimate (\ref{eq:bound}) on $G_i$ and shifting back, the bound 
(\ref{eq:Kestimate}) follows immediately. For the second part we note that 
\begin{equation*}
 \lim_{\sigma \rightarrow 1^+}\frac{1}{4\pi}\int_{-\infty}^{\infty}\limits 
 \rho\,h(\rho)\,\log(\sigma-1)\ \df \rho = 0 \ ,
\end{equation*}
thus we can add this term to $\Phi$. Then again shifting the integral 
by $-\imagunit \epsilon$ and using \cite[p. 44]{Magnus:1966} one obtains
\begin{equation*}
 \lim_{d(z',z) \rightarrow 0} \limits K_i(z',z)=
 \frac{1}{4\imagunit  \pi^2} \inte{-\imagunit \epsilon}\rho\,h(\rho)
 \left[\psi(\imagunit \rho)+\psi(\imagunit \rho +1)\right]\ \df \rho \ ,
 \quad i=1,4 \ ,
\end{equation*}
where $\psi(z)=\frac{\df}{\df z}\log\Gamma(z)$.
After substituting $\rho$ by $-\rho$ and using the fact that $h$ is even
the statement follows immediately.
\end{proof}
Having showed that $K$ is indeed a point-pair invariant,
we can introduce the auto\-morphic kernel
\begin{equation}
\label{eq:autokern}
 K_{\Gamma}(z',z):=\frac{1}{2}\sum_{\gamma\in \bar{\Gamma}}\limits 
 K(z',\gamma z)\chi(\gamma)J_{\gamma}(z,1)
\end{equation}
which by construction is in $\mathfrak{L}^2(\Gamma\backslash\mathbb{H}^2 ,
\Gamma\backslash\mathbb{H}^2,\chi)$. The corresponding Hilbert-Schmidt 
operator on $\mathfrak{L}^2(\Gamma,1,\chi)$ is called
$\mathrm{L}$.
\begin{lemma}
Any $\Psi \in \mathfrak{F}_{\rho}(\Gamma,1,\chi)$ is simultaneously an
eigenform of the Hilbert-Schmidt operator $\mathrm{L}$,
\begin{equation*}
 [\mathrm{L}\Psi](z')=\Lambda(\rho)\Psi(z') \  ,
\end{equation*}
where the eigenvalue $\Lambda$ satisfies the equation
\begin{equation}
\label{eq:lambda}
 \Lambda(\rho)+\Lambda(-\rho)=2h(\rho) \ .
\end{equation}
\end{lemma}
\begin{proof}
With
\begin{equation*}
 [\mathrm{L}\psi](z')=\frac{1}{2\pi \imagunit} \int_{F}\limits
 \left( \sum_{\gamma\in \bar{\Gamma}}\int_{-\infty}^{\infty}\limits h(\rho)\,
 G(z',\gamma z;\rho)\,\chi(\gamma)\,J_{\gamma}(z,1)\,\df \rho \right)
 \Psi(z)\ \df \mu(z)
\end{equation*}
a standard calculation yields
\begin{equation*}
 [\mathrm{L}\psi](z')=\frac{1}{\pi \imagunit} \inte{-\imagunit\beta}h(\rho')
 \left(\int_F\limits G_{\Gamma}(z',z;\rho')\,\Psi(z)\ \df \mu(z)\right)
 \df \rho' \ .
\end{equation*}
From this we read off the eigenvalue of $\mathrm{L}$ as
\begin{equation*}
 \Lambda(\rho)=\frac{1}{2\pi \imagunit}\int_{-\infty-\imagunit 
 \beta}^{\infty-\imagunit \beta}\limits\frac{h(\rho')}{\rho'-\rho}\ 
 \df \rho' \ ,
\end{equation*}
and this implies (\ref{eq:lambda}).
\end{proof}
As already noted in Section~\ref{section:sym}, the spectrum of the Dirac 
operator on a compact surface is real and discrete. Moreover, according
to Lemma~\ref{lem:chiral} the spectrum is symmetric with respect to zero.
We hence denote the non-negative eigenvalues (listed with their respective
multiplicities) by $0\leq\rho_0\leq\rho_1\leq\dots$, which include $N$ of 
the $2N$ zero modes. The Hilbert space 
$\mathfrak{L}^2(\Gamma,1,\chi)$ therefore has a basis 
of orthonormal eigenforms $\Psi_{n,\pm}$,
\begin{equation*}
 (\mathrm{D}\pm \rho_n)\Psi_{n,\pm}=0 \ .
\end{equation*}
Thus the automorphic kernel possesses a spectral expansion of the form
\begin{equation*}
 K_{\Gamma}(z',z)=\sum_{n=0}^{\infty}\limits\left(\Lambda(\rho_n)\,
 \Psi_{n,+}(z')\Psi^{\dagger}_{n,+}(z)+\Lambda(-\rho_n)\,\Psi_{n,-}(z')
 \Psi^{\dagger}_{n,-}(z)\right) \ ,
\end{equation*}
which immediately yields the spectral side of the desired trace formula.
\begin{lemma}
\label{korollar:lhs}
The Hilbert-Schmidt operator $\mathrm{L}$ has a finite trace, given by
\begin{equation*}
 \mathrm{tr}(\mathrm{L})=2\sum_{n=0}^{\infty}\limits h(\rho_n) \ .
\end{equation*}
\end{lemma}
Next we need to compute the geometric side of the trace formula, i.e.,
\begin{equation*}
 \mathrm{tr}(\mathrm{L}) =  \int_{F}\limits K_{\Gamma}(z,z)\ \df\mu(z)\ ,
\end{equation*}
with the representation (\ref{eq:autokern}) of the automorphic kernel. 
Following a standard procedure, this yields
\begin{equation}
\label{eq:tra1}
 \mathrm{tr}(\mathrm{L})=\frac{1}{2}\sum_{\{\gamma\}}\limits \chi(\gamma)
 \sum_{[g]\in\bar{Z}_{\gamma}\backslash\bar{\Gamma}}\mathrm{tr}
 \left(\,\int_{gF}\limits K(z,\gamma z)\,J_{\gamma}(z,1)\ \df\mu(z)
 \right) \ ,
\end{equation}
where $\{\gamma\}$ denotes the $\bar{\Gamma}$-conjugacy classes of 
$\gamma\in\bar{\Gamma}$ and $\bar{Z}_{\gamma}$ is the centraliser of $\gamma$
in $\bar{\Gamma}$. Next we use the natural pairing of the disjoint conjugacy 
classes $\{\gamma\}$ and $\{-\gamma\}$, which cancels the factor 
$\tfrac{1}{2}$. Within these pairs we choose the conjugacy classes with 
$\mathrm{tr}(\gamma)>0$. Moreover, it is well-known that
\begin{equation*}
 D_{\gamma}:=\bigcup_{[g]\in Z_{\gamma}\backslash\Gamma}g(F)
\end{equation*}
is a fundamental set for the centraliser $Z_{\gamma}\subset\Gamma$. Introducing
primitive hyperbolic elements $\gamma_p$ and their conjugacy classes,
one can rewrite (\ref{eq:tra1}) as
\begin{equation*}
\begin{split}
\label{eq:jj}
 \mathrm{tr}(L)=\ &\mathrm{tr}\left(\int_{F} K(z,z)\ \df \mu(z)\right)\\
 &+\sum_{\{\gamma_p\}}\sum_{n=1}^{\infty} \chi(\gamma_p^n)\,
 \mathrm{tr}\left(\int_{D_{\gamma_p}} K(z,\gamma_p^nz)\,J_{\gamma_p^n}(z,1)
 \ \df \mu(z)\right) \ .
\end{split}
\end{equation*}
Upon a conjugation with a matrix in $\mathrm{SL}(2,\mathbb{R})$ any hyperbolic 
element can be brought into the Jordan normal form
$\gamma=\left(\begin{smallmatrix}\ez^{\frac{l_{\gamma}}{2}}&0\\ 
0& \ez^{-\frac{l_{\gamma}}{2}}\end{smallmatrix}\right)$. A fundamental domain 
for the group generated by $\gamma$ is given by 
$\{z\in\mathbb{H}^2|1<y<\ez^{l_{\gamma}}\}$. Note that $J_{\gamma}(z,1)=1$, 
so all that remains to be done is evaluating integrals of the form
\begin{equation}
\label{eq:hyp}
 I(\gamma^n)=\mathrm{tr}\left(\int_1^{\ez^l}\limits\intel{} K(z,\gamma^n z)
 \ \df x\frac{\df y}{y^2} \right) \ .
\end{equation}
In order to calculate this integral we introduce
\begin{equation*}
 \tau=\cosh^2\frac{nl}{2}+\left(\frac{x}{y}\sinh\frac{nl}{2}\right)^2 \ ,
\end{equation*}
which allows us to perform the integration with respect to $y$.
Inserting the explicit expressions (\ref{eq:Hdiag}) and (\ref{eq:Hnondiag}) 
for the function $H$, the integral (\ref{eq:hyp}) reduces to 
\begin{multline*}
 I(\gamma^n)=\frac{\imagunit l_{\gamma}}{2 \pi^2}\coth\frac{nl_{\gamma}}
 {2}\intel{-\imagunit \epsilon}\rho\,h(\rho) \times  \\ \times
 \int_{\cosh^2\frac{nl_{\gamma}}{2}}^{\infty}\limits 
 \tau^{-1-\imagunit \rho}\left(\tau-\cosh^2\frac{nl_{\gamma}}{2}
 \right)^{-\frac{1}{2}}\sum_{m=0}^{\infty}\limits
 \frac{\Gamma(\imagunit \rho+m)\Gamma(\imagunit \rho + m +1)}
 {\Gamma(2\imagunit \rho +m +1)m!}\tau^{-m}\df\tau \df\rho \ .
\end{multline*}
Interchanging the order of integration and summation and employing the
relation
\begin{equation*}
 \left(\frac{1}{2\cosh{(\frac{a}{2})}}\right)^{-2q} \sum_{m=0}^{\infty}\limits
 \frac{2q\Gamma(2q+2m)}{\Gamma(2q+m+1) m!} \left(
 \frac{1}{2\cosh{(\frac{a}{2})}}\right)^{-2m} = \ez^{-qa} \ ,
\end{equation*}
which follows from \cite[1.114]{Gradshteyn:1965} for $a>0$ and 
$\mathrm{Re}(q)>0$, finally yields
\begin{equation}
\label{eq:hypp}
 I(\gamma^n)=\frac{l_{\gamma}}{\sinh\frac{nl_{\gamma}}{2}}
 \frac{1}{2\pi}\intel{-\imagunit \epsilon}h(\rho)\,
 \ez^{-\imagunit\rho nl_{\gamma}}\ \df \rho  =
 \frac{l_{\gamma}g(nl_{\gamma})}{\sinh(\tfrac{nl_{\gamma}}{2})} \ .
\end{equation}
Putting together Lemma~\ref{lemma:id}, Lemma~\ref{korollar:lhs}
and equation (\ref{eq:hypp}) then leads to the desired trace formula.
\begin{thm}[Selberg trace formula for the Dirac operator]
Let $\Gamma\subset \psl$ be a strictly hyperbolic Fuchsian group with 
fundamental domain $F$ of area $A(F)$ and fix a multiplier system $\chi$
of weight one. Moreover, let $\{\rho_n\}_{n=0}^{\infty}$ be the non-negative 
eigenvalues of the Dirac operator $\mathrm{D}$ on 
$\mathfrak{L}^2(\Gamma,1,\chi)$, including half of the zero-modes. Then, 
for any admissible test function $h$, see Definition~\ref{defn:h}, the 
following trace formula holds,
\begin{equation}
\label{eq:spurformel}
 \sum_{m=0}^{\infty}\limits h(\rho_m)=
 \frac{A(F)}{4\pi}\int_{-\infty}^{\infty}\limits \rho\,h(\rho)\,
 \coth(\pi\rho)\ \df\rho
 +\sum_{\{\gamma_p\}}\limits\sum_{n=1}^{\infty}\limits \chi(\gamma_p^n)\,
 \frac{l_{\gamma_p}\,g(nl_{\gamma_p})}{2\sinh(\frac{nl_{\gamma_p}}{2})}\ .
\end{equation}
\end{thm}
In accordance with Proposition~\ref{satz:auto}, this trace formula is 
identical to the one for the Maa\ss -Laplacian $-\Delta_1$ on the same 
surface and with the same multiplier system. We stress, however, a 
difference in the interpretation of the geometric side. Both expressions
can be viewed as sums over the closed geodesics on 
$\Gamma\backslash\mathbb{H}^2$, weighted with the factors $\chi(\gamma_p^n)$.
For the Maa\ss -Laplacian these factors stem from the non-vanishing magnetic 
fluxes that are necessarily present (see, e.g., \cite{Comtet:1993}), whereas 
for the Dirac operator with weight $k=1$ there is no magnetic field involved. 
Here the factors $\chi(\gamma_p^n)$ reflect the fact that the classical 
dynamical system associated with the quantum dynamics generated by the Dirac 
operator is not the geodesic flow, but its associated frame flow, see also 
\cite{Jakobson:2006}. This is analogous to Dirac operators on $\mathbb{R}^3$ 
\cite{Bolte:1999}.

The frame flow (see, e.g., \cite{Brin:1980}) is a flow in the frame bundle
$FM$ over a Riemannian manifold $M$ consisting of a parallel transport of
oriented orthonormal frames along geodesics. To be precise, let
$\{e_1(p),\dots,e_n(p)\}$ be an orthonormal basis of $T_p M$. Then this
frame is transported along the geodesic determined by $e_1$ with the
Levi-Civita connection. This flow can be lifted into the spin structure
by assigning an element $g(p)\in\mathrm{Spin}(n)$ to every point along the 
geodesic. This may also be done in the spin representation yielding
$\rho(g(p))$. In the present case, where $M=\Gamma\backslash\mathbb{H}^2$, 
$\rho(g(z))$ must equal $\rho(g(\gamma z))$ for every $\gamma\in\bar{\Gamma}$.
However, in the induced (trivial) bundle over the hyperbolic plane the
transformation property
\begin{equation*}
 \rho(g(\gamma z))=\chi(\gamma)\,J_{\gamma}(z,1)\,\rho(g(z))
\end{equation*}
applies, see (\ref{eq:factaut}) and (\ref{eq:Diracauto}). The factors
$\chi(\gamma_p^n)$ in equation (\ref{eq:spurformel}) therefore reflect
the spin structure. Hence, although on two-dimensional manifolds the frame 
flow is not much different from the geodesic flow, only the latter is
a natural dynamics that can be lifted into the spin structure. For
$n\geq 3$ the situation will be different since in such a case a frame flow 
yields interesting classical dynamics beyond the geodesic flow.
\section{Some Applications}
\label{section:applications}
A first immediate application of the trace formula, which can be proved
in the standard way (see, e.g., \cite{Hejhal:1976}), concerns the asymptotic 
distribution of the eigenvalues.
\begin{satz}[Weyl's law]
Let $N(\rho)$ be the number of non-negative eigenvalues of $D$ which are 
smaller than $\rho$. Then
\begin{equation*}
 N(\rho)\sim \frac{A(F)}{4\pi}\rho^2,\quad \rho \rightarrow \infty.
\end{equation*}
\end{satz}
Another application consists of determining properties of the related 
Selberg zeta function. If $\mathrm{Re}(s),\,\mathrm{Re}(\sigma)>1$ the
function
\begin{equation*}
 h(\rho):=\frac{1}{\rho^2+(s-\frac{1}{2})^2}
 -\frac{1}{\rho^2+(\sigma-\frac{1}{2})^2}
\end{equation*}
satisfies the criteria of Definition~\ref{defn:h} to serve as an admissible 
test function. In the trace formula (\ref{eq:spurformel}) it leads to a 
relation for the trace of a regularised resolvent,
\begin{equation*}
\begin{split}
 \sum_{m=0}^{\infty}\limits \left(   \frac{1}{\rho_m^2+(s-\frac{1}{2})^2}
 -\frac{1}{\rho_m^2+(\sigma-\frac{1}{2})^2}\right) 
 =& -\frac{A(F)}{2\pi} \left(\psi(s-\tfrac{1}{2})-\psi(\sigma-\tfrac{1}{2})
    \right) \\
  &+\frac{A(F)}{4 \pi}\left(\frac{1}{\sigma-\frac{1}{2}}  
   -\frac{1}{s-\frac{1}{2}}  \right) \\
  &+\frac{1}{2s-1}\frac{Z'(s)}{Z(s)}- \frac{1}{2\sigma-1}
   \frac{Z'(\sigma)}{Z(\sigma)}\ .
\end{split}
\end{equation*}
Here $Z(s)$ is Selberg's zeta function which is defined by
\begin{equation*}
 Z(s):=\prod_{\{\gamma_p\}}\limits \prod_{k=0}^{\infty}\limits
 (1-\chi(\gamma_p)\,\ez^{-l_{\gamma_p}(k+s)})\ ,\quad 
 \mathrm{Re}(s)>1\ .
\end{equation*}
Now proceeding along the lines of \cite{Steiner:1987,Sarnak:1987}
we find an analytic continuation of $Z(s)$ into the entire complex plane.
\begin{satz}
The Selberg zeta function for the Dirac operator is an entire analytic
function. Moreover, it can be represented as
\begin{equation}
\label{eq:zeta}
\begin{split}
 Z(s)= &\frac{Z^{(2N)}(\tfrac{1}{2})}{(2N)!}  \left(s-\frac{1}{2}\right)^{2N} 
        \ez^{(s-\frac{1}{2})^2\gamma_D}\ez^{(s-\frac{1}{2})\frac{A(F)}{2\pi}}
        \times\\ 
       &\times\left[(2\pi)^{-(s-\frac{1}{2})} \ez^{(s^2-\frac{1}{4})}
        G^2(s+\tfrac{1}{2})\right]^{\frac{A(F)}{2\pi}}
        \prod_{m=N}^{\infty}\limits \left[\left(1+  \frac{(s-\frac{1}{2})^2} 
        {\rho_m^2}\right) \ez^{-\frac{(s-\frac{1}{2})^2}{\rho_m^2}}\right]\ ,
\end{split}
\end{equation}
where $G$ is Barnes' double $\Gamma$-function (see \cite{Barnes:1899})
and $2N$ denotes the number of zero modes of $\mathrm{D}$. The trivial zeroes 
of $Z(s)$ are given by $s=-\tfrac{1}{2}-n$ with multiplicities 
$(2n)^{\frac{A(F)}{2\pi}}$, $n\in\mathbb{N}$. The non-trivial zeroes are  
$s=\tfrac{1}{2}\pm\imagunit \rho_m$, with the same multiplicities as the 
eigenvalues $\rho_m$ of $\mathrm{D}$.
\end{satz}
In (\ref{eq:zeta}) the constant $\gamma_D$ is a generalised Euler constant
as introduced in \cite{Steiner:1987} and $Z^{(2N)}$ denotes a derivative
of the zeta function of order $2N$. Moreover, due to the Gau\ss -Bonnet
theorem, $\frac{A(F)}{2\pi}=2(g-1)$ is a positive integer, where $g$ is
the genus of the surface $\Gamma\backslash\mathbb{H}^2$.
{\small
\bibliographystyle{amsalpha}
\bibliography{diplom}
}

\end{document}